\documentstyle{article}
\begin{document}
\leftline{\sl Text of an invited talk given at}
\leftline{\sl The 8th International Conference on Recent Progress in
              Many-body Physics}
\leftline{\sl 22-27 August 1994, Schlo\ss Seggau, Styria, Austria}

\vspace{2 \baselineskip}
\begin{center}
  {\bf A NONPERTURBATIVE MICROSCOPIC THEORY OF\\
  HAMILTONIAN LATTICE GAUGE SYSTEMS}\\
  \vspace{\baselineskip}
  {R.F. Bishop, N.J. Davidson, and Y. Xian}\\
  \vspace{\baselineskip}
  {\it Department of Mathematics, UMIST\\
  (University of Manchester Institute of Science and Technology)\\
  P.O. Box 88, Manchester M60 1QD, UK}
\end{center}

\vspace{\baselineskip}
A review is given of our recent application of a systematic
microscopic formulation of quantum many-body theory, namely the
coupled-cluster method (CCM), to Hamiltonian $U(1)$ lattice gauge
models in the pure gauge sector. It is emphasized that our CCM results
represent a natural extension or resummation of the results from the
strong-coupling perturbation theory.

\section{INTRODUCTION}

Lattice gauge field theory was first developed by Wilson \cite{wils}
in Euclidean space-time to tackle the problem of quark confinement for
the strong interaction. Independently, the equivalent Hamiltonian
models were formulated by Kogut and Susskind \cite{kosu}. The lattice
supplies an ultra-violet cut-off which regularizes the divergency
often encountered in continuum field theory. One of the key advantages
of lattice gauge theory clearly lies in the fact that the confining
strong-coupling limit provides a natural basis from which one can
apply such techniques as perturbation theory and other many-body
theory approximations. The fact that the physical continuum limit is
achieved in the weak-coupling limit provides a stringent test for any
technique applied to lattice gauge theory.

There is a general theorem which states that all lattice gauge models
possess a nonzero confining region in which the strong-coupling
perturbation theory is valid \cite{osse}. In other words, the
strong-coupling perturbation series of all lattice gauge models have a
finite radius of convergence. One challenge in lattice gauge theory is
to extend the strong-coupling results to the weak-coupling regime.
Methods based on Pad\'e approximants and similar techniques are often
used for this purpose \cite{haois}.  However, this rather {\it ad hoc}
approach requires a prior knowledge of the weak-coupling limit. Among
many other techniques, including finite size calculations
\cite{finite}, renormalization group methods \cite{reno},
$t$-expansion techniques \cite{texp}, and loop calculus \cite{loop}
etc., the numerical Monte Carlo simulations
\cite{MC} seem provide the most reliable results, although the
method is computationally intensive in practice. Recently, several
attempts have been made to apply powerful many-body theories to
Hamiltonian lattice gauge systems. Two such applications include the
method of correlated basis functions (CBF) \cite{CBF} and the
coupled-cluster method (CCM) \cite{CCM,wats}, both of which provide
intrinsically nonperturbative results.

In this article, we review our recent progress in the application of
the CCM to the vacuum state of the $U(1)$ lattice gauge theory in
$1+1$, $2+1$ and $3+1$ dimensions (referred to as 1D, 2D, and 3D
respectively). The 1D model consists of a linear array of plaquettes,
while the 2D and 3D models are based on the square and cubic lattices
respectively. In particular, we have formulated the lattice gauge
Hamiltonian in terms of a many-body theory and applied several
well-tested approximation schemes within the framework of the CCM.
These approximation schemes of the CCM have been developed by us for
quantum spin lattice systems and have met with considerable success
\cite{CCMspin1}. Not only are they able to produce results
for quantum spin lattice models with accuracy comparable to that of
the best Monte Carlo calculations, but they also enable us to study
the possible quantum phase transitions of the systems in a systematic
and unbiased manner \cite{CCMspin2}. This second ability of the CCM
may prove especially significant here since lattice gauge systems may
experience a deconfining phase transition as the coupling parameter
varies from strong to weak.  We notice that the 3D $U(1)$ gauge
lattice model must recover a deconfined continuum QED in the weak
coupling limit.  However, it is widely believed that the confining
phase persists for all couplings for the $U(1)$ models in 1D and 2D,
and for the $SU(2)$ and $SU(3)$ models in less than four spatial
dimensions \cite{momu}.

The rest of our article is organized as follows. In Sec. 2 we first
discuss the number of independent degrees of freedom for the $U(1)$
lattice models in the pure gauge sector, and then transform the gauge
invariant Hamiltonian into a many-body Hamiltonian. We present the
results for the ground-state energy as a function of the coupling
parameter for the $U(1)$ models in 1D and 2D in Sec. 3, and the
results of the 3D model in Sec. 4. We conclude our article with a
discussion in Sec. 5.

\section{THE U(1) MODELS AND THEIR DEGREES OF FREEDOM}

In lattice gauge models, the physical fields are defined on the
directed links $\{l\}$ of the lattice. In particular, the Abelian
$U(1)$ lattice Hamiltonian after suitable scaling can be written as
\cite{kosu,CCM}
\begin{equation}
  H = \sum_{l=1}^{N_l}{1\over2}E_l^2
       +\lambda\sum_{p=1}^{N_p}(1-\cos B_p),
\end{equation}
where the link index $l$ runs over all $N_l$ links of the lattice, the
plaquette index $p$ over all $N_p$ elementary lattice plaquettes, and
$\lambda$ is the coupling constant, with $\lambda=0$ being referred to
as the strong-coupling limit and $\lambda\rightarrow\infty$ as the
weak-coupling limit. Clearly, we shall be interested in the bulk
(thermodynamic) limit, where $N_l, N_p \rightarrow \infty$. If $N_s$
is the number of lattice sites, it is easy to see that in the bulk
limit we have $N_l = 3N_p, N_s = 2N_p$ in 1D; $N_l = 2N_p, N_s = N_p$
in 2D; and $N_l = N_p, N_s = N_p/3$ in 3D. The electric field $E_l$
is defined on the link $l$, while the magnetic field $B_p$ is a
plaquette variable, given by the lattice curl of the link-variable
vector potential $A_l$ as
\begin{equation}
 B_p \equiv A_{l_1} + A_{l_2} - A_{l_3} - A_{l_4},
\end{equation}
with the four links, $l_1,l_2,l_3$ and $l_4$, enclosing an elementary
square plaquette $p$ in the counter-clockwise direction. The direction
of the magnetic field ${\bf B}_p$ can be defined by the right-hand
rule around the plaquette. The quantization of the fields is given by
the commutator,
\begin{equation}
  [A_l, E_{l'}] = i\delta_{ll'}.
\end{equation}
If we choose the representation, $E_l = -i \partial /\partial A_l$,
the Hamiltonian of Eq. (1) becomes
\begin{equation}
  H = -\sum_{l=1}^{N_l}{1\over2}{\partial^2\over\partial A_l^2}
       +\lambda\sum_{p=1}^{N_p}(1-\cos B_p),
\end{equation}
where the compact variable $A_l$ is restricted to the region $-\pi <
A_l\le\pi$.  The inner product between states $\vert\Psi(\{A_l\})
\rangle$ and $\vert\tilde\Psi(\{A_l\})\rangle$ is defined as:
\begin{equation}
  \langle\tilde\Psi\vert\Psi\rangle_A =
  \prod_{l=1}^{N_l}\left(\int^\pi_{-\pi}{dA_l\over2\pi}\right)\,
  \tilde\Psi^*(\{A_l\})\Psi(\{A_l\}).
\end{equation}

It is useful to denote each link $l$ by both a lattice site vector
$\bf n$ and an index $\alpha$ indicating direction, $\alpha=\pm x,\pm
y,\pm z$, so that $E_l \equiv E_\alpha({\bf n})$, for example. By
definition, one has $E_{-x}({\bf n}) = - E_x({\bf n}-\hat x)$, etc.,
where $\hat x$ is a unit lattice vector in the $+x$-direction.
Similar definitions hold for the vector potential $A_\alpha({\bf n})$.
The lattice divergence of the electric field on a site $\bf n$ can now
be written as
\begin{equation}
  (\nabla\cdot{\bf E})({\bf n}) = \sum_{\alpha} E_\alpha({\bf n});
\end{equation}
where the summation is over $\alpha=\pm x,\pm y,\pm z$. A gauge
transformation of any operator, such as the vector potential
$A_x({\bf n})$, is given by
\begin{equation}
  \exp\left[i\sum_{\bf m}^{N_s}\phi({\bf m})
   (\nabla\cdot{\bf E})({\bf m})\right]
  A_x({\bf n})\exp\left[-i\sum_{\bf m}^{N_s}
  \phi({\bf m})(\nabla\cdot{\bf E})({\bf m})\right]
  = A_x({\bf n}) + \phi({\bf n}) - \phi({\bf n}+\hat x),
\end{equation}
where $N_s$ is the total number of lattice sites and $\phi({\bf m})$
is an arbitrary gauge function. Clearly, the plaquette variable $B_p$
is invariant under this gauge transformation according to the
definition of Eq. (2). It is also easy to show that the Hamiltonian of
Eq. (4) is invariant under this gauge transformation, as expected.

One can also define the divergence of the plaquette variable $B_p$ on
a lattice site ${\bf n}$. Clearly, this divergence is zero in 1D and
2D because the plaquette direction (i.e., unit vector perpendicular to
the plaquette with the right-hand rule) is a constant. In 3D, the
plaquette direction varies from plaquette to plaquette. We associate
an elementary cube with each lattice site ${\bf n}$, with ${\bf n}$ at
the origin of the cube in Cartesian coordinates, and denote the six
plaquette variables as $B_\beta({\bf n})$ with plaquette direction
$\beta = \pm x,\pm y,\pm z$, where for example, $\beta = z$
represents the bottom plaquette of the cube, $\beta=-z$ the top
plaquette of the cube, etc. For later purposes, we refer to the three
plaquettes $B_\beta({\bf n})$ with $\beta = x,y,z$ as positive
plaquettes with respect to the cube at $\bf n$, and the other three
with $\beta = -x,-y,-z$ as negative. By definition, for the negative
plaquette, one has $B_{-z}({\bf n}) = - B_z({\bf n}+\hat z)$, etc.
The divergence of the plaquette variable at lattice site $\bf n$ can
then be clearly written as
\begin{equation}
  (\nabla\cdot{\bf B})({\bf n}) = - \sum_\beta  B_\beta({\bf n}),
\end{equation}
with summation over $\beta = \pm x,\pm y,\pm z$. We notice that upon
substitution of Eq. (2), Eq. (8) yields constant zero, as required by
the Bianchi identity, namely
\begin{equation}
  (\nabla\cdot{\bf B})({\bf n}) = 0, \ \ \ \ \forall {\bf n}.
\end{equation}

We now discuss the number of independent degrees of freedom. Since we
are working with the pure gauge Hamiltonians, we restrict ourselves to
the gauge-invariant (vacuum) sector of Hilbert space. We therefore
require that any state $\vert\Psi\rangle$ satisfies
\begin{equation}
  (\nabla\cdot{\bf E})({\bf n}) \vert\Psi\rangle = 0,  \ \ \ \
       \forall {\bf n}.
\end{equation}
This imposes $N_s$ restrictions on the $N_l$ vector potential
variables $\{A_l\}$, where $N_s$ is the number of lattice sites.
Therefore, the number of independent variables for $U(1)$ lattice
models in the vacuum sector is reduced to $N = N_l - N_s$. We also
note that if the wavefunction $\Psi$ is written as a function of
plaquette variables, namely $\Psi = \Psi(\{B_p\})$, Eq. (10) is
satisfied.

It is easy to see that for the infinite lattice in 1D and 2D, $N=N_p$,
namely, the number of independent degrees of freedom is equal to the
number of plaquette variables. Therefore, it is proper and convenient
to employ the plaquette variables $\{B_p\}$ for the 1D and 2D models.
The corresponding inner product between two states $\vert\Psi(\{B_p\}
\rangle$ and $\vert\tilde\Psi(\{B_p\})\rangle$ is then defined by
integrals over all plaquette variables,
\begin{equation}
  \langle\tilde\Psi\vert\Psi\rangle_B =
  \prod_{p=1}^{N_p}\left(\int^\pi_{-\pi}{dB_p\over2\pi}\right)\,
  \tilde\Psi^*(\{B_p\})\Psi(\{B_p\}).
\end{equation}

For the infinite 3D lattice model, however, one has that $N = 2N_p/3$.
If one is to employ the plaquette variables $\{B_p\}$ in taking the
inner products as discussed above for the 1D and 2D models, one still
has to satisfy the $N_s\ (= N_p/3)$
geometrical constraints of the Bianchi
identity, Eq. (9). In general, these restrictions are quite difficult
to apply. It is therefore more convenient to employ the $N_l\ (= N_p)$
link variables $\{A_l\}$ for the 3D model when taking the inner
products, as defined by Eq. (5).  The gauge invariance constraint of
Eq. (10) are then automatically satisfied so long as the wavefunctions
are completely expressible in terms of plaquette variables $\{B_p\}$.
Since we are dealing with compact lattice gauge theory (i.e., $-\pi <
A_l \le \pi$), the redundant degrees of freedom in $\{A_l\}$ have no
effect on evaluating expectation values with normalized wavefunctions.

The conclusion of the above discussion is that when taking inner
products, we shall employ plaquette variables $\{B_p\}$ for the 1D and
2D models, and employ link variables $\{A_l\}$ for the 3D case; but
the wavefunctions of all models should always be expressible in terms
of the plaquette variables alone. Therefore, it is convenient to
transform the Hamiltonian of Eq. (4) into a form in which only
plaquette variables appear. By using the linear relation of Eq. (2),
this can be easily done. We thus derive
\begin{equation}
 H=\sum_{p=1}^{N_p} \left[ -2{\partial^2\over\partial B_p^2}
      +\lambda (1+\cos B_p)\right]
      +{1\over2}\sum_p^{N_p}\sum_\rho^z(-1)^\rho {\partial^2 \over
      \partial B_p\partial B_{p+\rho}},
\end{equation}
where $\rho$ is the nearest-neighbour plaquette index, the summation
over it runs over all $z$ nearest-neighbours plaquettes, and where we
have employed the notation
\begin{equation}
   (-1)^\rho = \left\{
  \begin{array}{ll}
    1, &\rho\in\rho_\parallel;\\
    (-1)^{\rho_\perp},&\rho\in\rho_\perp,
  \end{array} \right.
\end{equation}
and
\begin{equation}
 (-1)^{\rho_\perp} = \left\{
 \begin{array}{ll}
   1, &{\rm if}\ p\ {\rm and}\ p+\rho_\perp\ {\rm denote\ both\
   positive\ or\ both\ negative\ plaquettes};\\
  -1, &{\rm otherwise.}
 \end{array}         \right.
\end{equation}
In Eqs. (13) and (14), $\rho_\parallel$ and $\rho_\perp$ denote
nearest-neighbour parallel and perpendicular plaquette indices
respectively, and the positive or negative plaquettes have been
defined in the paragraph before Eq. (8).

\section{GROUND-STATE ENERGY FOR THE 1D AND 2D MODELS}

As discussed in Sec. 2, the proper variables for the 1D and 2D $U(1)$
models are the plaquette variables $\{B_p\}$. The Hamiltonian of Eq.
(12), for the 1D and 2D cases, reduces to
\begin{equation}
 H=\sum_{p=1}^{N_p}\biggl[ -2{\partial^2\over\partial B_p^2}
+\lambda
 (1-\cos B_p)\biggr] +{1\over2}\sum_{p=1}^{N_p}
 \sum_\rho^z{\partial^2\over \partial
 B_p\partial B_{p+\rho}};\ \ \ \ -\pi \le B_p \le \pi,
\end{equation}
where we have dropped the parallel symbol, $\rho = \rho_\parallel$,
and $z = 2,4$ for the 1D and 2D cases respectively.

The details of our calculations for Eq. (15) have been published
elsewhere \cite{CCM}. In particular, we first consider the independent
plaquette Hamiltonian in the strong-coupling limit ($\lambda =0$),
$H_0 = -2\sum_p d^2/dB^2_p$, which has two sets of eigenstates, namely
$\{\cos mB;\ m=0,1,2,...\}$ with even parity and $\{\sin mB;\ m =
1,2,...\}$ with odd parity. The corresponding ground state is clearly
a constant, which is referred to as the electric vacuum in the
literature. We take this electric vacuum state as our CCM model state
$\vert\Phi\rangle$. Hence we have, $\vert\Phi\rangle = C=1$. The
many-plaquette exact ground state $\vert\Psi_g\rangle$ of the full
Hamiltonian is, according to the CCM ansatz, written as
\begin{equation}
  \vert\Psi_g\rangle = {\rm e}^S\vert\Phi\rangle, \ \ \ \
  S=\sum_{k=1}^{N_p} S_k,
\end{equation}
where the correlation operator $S$ is partitioned into $k$-plaquette
operators $\{S_k\}$. For example, the one-plaquette operator is
defined as
\begin{equation}
 S_1 = \sum_{n=1}^\infty\sum_{p=1}^{N_p} {\cal S}_p(n)\cos nB_p;
\end{equation}
and the two-plaquette operator consists of two terms,
\begin{eqnarray}
 S_2 = {1\over2!}\sum_{n_1,n_2=1}^\infty \mathop{{\sum}'}_{p_1,
  p_2=1}^{N_p}&& \bigl [{\cal S}_{p_1p_2}^{(1)}(n_1,n_2)
  \cos n_1B_{p_1}\cos n_2B_{p_2} \nonumber \\
 &&+{\cal S}_{p_1p_2}^{(2)}(n_1,n_2)\sin n_1B_{p_1}\sin
n_2B_{p_2}\bigr],
\end{eqnarray}
where the prime on the summation excludes the terms with $p_1=p_2$.
We note that the many-plaquette correlation operators $S_k$ have a
close relation to the usual Wilson loops \cite{wils,kosu}. For
example, one can write $2\cos B_1\cos B_2 = \cos(B_1+B_2) +
\cos(B_1-B_2)$, which corresponds to the following relation for the
Wilson loops:
\begin{figure}[h]
\vspace{2.4cm}\hspace{3cm}\special{fig0.ps}
\vspace{-1cm}
\end{figure}

\noindent
Our parametrization exemplified by Eqs. (17) and (18) is clearly
complete. It is also particularly useful in view of the orthonormality
of the basis. However, for the 3D model, since we have to employ the
link variables when taking inner products, this orthonormality is in
some sense lost. We shall discuss this point in the next section.

{}From the Schr\"odinger ground-state equation, $H\vert\Psi_g\rangle =
E_g\vert\Psi_g\rangle$, or ${\rm e}^{-S}H{\rm e}^S\vert\Phi\rangle =
E_g\vert\Phi\rangle$, we obtain the equations for the ground-state
energy $E_g$ and the correlation coefficients $\{{\cal S}_p, {\cal
S}_{p_1p2}, \cdots\}$ by taking proper projections. In particular, the
energy equation and the one-plaquette equations can be written
together as
\begin{equation}
  \langle\Phi\vert\cos nB_p{\rm e}^{-S}H{\rm e}^S
   \vert\Phi\rangle_B =  E_g\delta_{n,0};\ \ \ \ n=0,1,2,...,
\end{equation}
and the two-plaquette equations consist of two sets of inner products,
\begin{eqnarray}
  \langle\Phi\vert\cos n_1B_{p_1}\cos n_2B_{p_2}{\rm e}^{-S}H{\rm
   e}^S \vert\Phi\rangle_B &=& 0,\\
 \langle\Phi\vert\sin n_1B_{p_1}\sin n_2B_{p_2}
  {\rm e}^{-S}H{\rm e}^S\vert\Phi\rangle_B  &=& 0,
\end{eqnarray}
where $n_1,n_2 =1,2,...$ and $p_1\ne p_2$ as before. The higher-order
equations can be written down in a similar fashion. In Eqs.
(19)-(21), the notation $\langle\cdots\rangle_B$ implies that the
inner products are integrals over all plaquette variables $\{B_p\}$,
as defined by Eq.  (11).

As usual, one needs to employ a truncation scheme for the correlation
operator $S$. We first consider the SUB1 scheme, in which one sets
$S_k=0$ for all $k > 1$. After an extension of the definition for the
one-plaquette coefficients $\{{\cal S}_p(n)\}$ to include the negative
modes (negative $n$), and taking advantage of the lattice
translational invariance to introduce the definition, $a_m\equiv
m{\cal S}_p(m)$, Eq.~(19) can be readily written as
\begin{equation}
   \left({E_g\over N_p}-\lambda\right)\delta_{m,0}
   +{1\over2}\lambda(\delta_{m,1}+\delta_{m,-1}) - ma_m
   -{1\over2}\sum_{n=-\infty}^\infty a_na_{m-n} =0,
\end{equation}
where $m$ may be any integer. We note that the energy equation is
given by setting $m=0$. Equation (22) can in fact be transformed to
the well-known Mathieu equation corresponding to the single-body
Schr\"odinger equation with the one-plaquette Hamiltonian given by the
first term of Eq.~(15) \cite{CCM}. We solve these SUB1 equations
numerically by a hierarchical sub-truncation scheme, the so-called
SUB1($n$) scheme in which one retains at the $n$th level of
approximation only those coefficients $a_m$ with $\vert m\vert \le n$,
and sets the remainder with $\vert m\vert >n$ to zero. For example, in
the SUB1(1) scheme, $a_1$ is the only retained coefficient. The
solution is trivially obtained from Eq.~(22) as
\begin{equation}
 a_1= {\lambda\over2}, \ \ \ \  {E_g\over N_p}= \lambda
    -{1\over4}\lambda^2; \ \ \ \ {\rm SUB1(1)}.
\end{equation}
This SUB1(1) result is in fact identical to the result obtained from
second-order perturbation theory about the strong-coupling
($\lambda\rightarrow0$) limit. However, the subsequent SUB1($n$)
approximations with $n>1$ give results far superior to those of
perturbation theory. A detailed discussion has been given in
Ref.~[11].

We next discuss the two-plaquette approximation, i.e., the SUB2 scheme
in which one makes the substitution $S\rightarrow S_{\rm
SUB2}=S_1+S_2$.  As defined by Eq.~(18), there are two sets of
two-plaquette coefficients which are determined by Eqs.~(20) and (21)
respectively. Together with the one-plaquette equations discussed
above, one has three sets of coupled equations. Since the complete
SUB2 approximation is very ambitious as a first attempt to include
correlations, we employ instead the local approximation developed by
us for the spin-lattice models, namely the LSUB$m$ scheme.
We consider just the LSUB2 scheme which includes only
nearest-neighbour plaquette correlations. Similar to the SUB1($n$)
scheme discussed above, we may also introduce the so-called LSUB2($n$)
sub-truncation scheme in terms of the number of modes $\{n_k\}$ kept
in the sums in Eqs. (17) and (18) by ignoring those terms in the LSUB2
correlation operator with $\sum_k n_k >n$. For example, the LSUB2(1)
scheme is identical to the SUB1(1) scheme considered previously in
which only a single coefficient $a_1$ is retained. In the LSUB2(2)
scheme, however, four coefficients are retained, two of them from the
one-plaquette correlations, the other two from the two-plaquette
correlations.

\begin{figure}
\vspace{8.4cm}
\hspace{1cm} \special{fig1.ps}
\vspace{-1.3cm}
\caption{Ground-state energy per plaquette for the 2D $U(1)$ model
on the square lattice in the LSUB2($n$) scheme. Also shown are the
full SUB1 results and the results from the $n$th-order
strong-coupling perturbation series, PT$n$.}
\vspace{\baselineskip}
\end{figure}

We notice that in the strong-coupling limit ($\lambda
\rightarrow 0$) the LSUB2(2) approximation exactly reproduces the
results of the corresponding perturbation series up to the fourth
order,
% \buildrel{\lambda\rightarrow0}\under
%  \def\normalbaselines{\baselineskip24pt
%     \lineskip3pt\lineskiplimit3pt}
\begin{equation}
   {E_g\over N_p}\rightarrow \left\{
   \begin{array}{ll}
    \lambda-{1\over4}\lambda^2+{89\over3840}\lambda^4+{\cal
   O}(\lambda^6), &{\rm 1D};\\
    \lambda-{1\over4}\lambda^2+{73\over3840}\lambda^4+{\cal
   O}(\lambda^6), &{\rm 2D}.
   \end{array} \right.
\end{equation}

The results for the ground-state energy in the LSUB2($n$) scheme up to
$n=10$ are shown as functions of $\lambda$ in Table~1 and 2 for the 1D
and 2D models respectively, together with some results of $n$th-order
strong-coupling perturbation theory, denoted as PT$n$(S). We note that
we have taken this opportunity to correct some minor errors in the
values cited previously in Ref. [11]. We also show the results of the
2D case in Fig.~1. The corresponding 1D curves behave similarly. In
Table~2 we have also included the results from the method of
correlated basis functions (CBF) \cite{CBF}, from an analytical
continuation of the strong-coupling perturbation series (HOZ)
\cite{haois}, and from the $t$-expansion calculation of Morningstar
\cite{texp}. Our LSUB2(10) results are in good agreement with them.
One sees clearly in Fig.~1 that our LSUB2($n$) results quickly
converge as $n$ increases. It is also clear that the strong-coupling
perturbation series gives very poor results for $\lambda \ge 1.5$, a
value which seems to be a good estimate for its radius of convergence.
Much work in modern quantum field theory goes into attempts to
continue analytically such perturbation series as Eq.~(24) outside
their natural boundaries.  A typical recent such attempt \cite{haois}
for the 2D $U(1)$ model starts from the strong-coupling perturbation
series of Eq.~(24), utilizing the known coefficients up to ${\cal
O}(\lambda^{18})$ as input to generalized Pad\'e approximants. The
results of this approximation are shown in Table 2 where they are
labelled as HOZ. We should emphasize that our own LSUB2($n$)
approximations themselves represent a natural extension of
perturbation theory. They may be contrasted with the rather {\it ad
hoc} approaches based on Pad\'e and other resummation techniques,
which usually find it difficult to approach the weak-coupling limit
with the correct asymptotic form unless this is built in from the
start.

\begin{table*}[t]
\setlength{\tabcolsep}{2.3mm}
\newlength{\digitwidth} \settowidth{\digitwidth}{\rm 0}
\caption{Ground-state energy per plaquette at several values of
$\lambda$ for the 1D U(1) model. Shown are the results from the CCM
LSUB2($n$) calculations, and from the strong- and weak-coupling
expansion series, denoted as PT$m$(S) ($m$th order) and PT(W)
respectively.}
\begin{tabular*}{\textwidth}{llllllllll}\hline\noalign{\smallskip}
            &\multicolumn{9}{c}{$\lambda$}\\
            \noalign{\smallskip}\cline{2-10}\noalign{\smallskip}
\multicolumn{1}{c}{Method}
            &\multicolumn{1}{c}{0.5}&\multicolumn{1}{c}{1}
            &\multicolumn{1}{c}{2}&\multicolumn{1}{c}{3}
            &\multicolumn{1}{c}{4}&\multicolumn{1}{c}{5}
            &\multicolumn{1}{c}{6}&\multicolumn{1}{c}{8}
            &\multicolumn{1}{c}{10}\\
            \noalign{\smallskip}\hline\noalign{\smallskip}
SUB1   &0.4391&0.7724&1.2430&1.5828&1.8597&2.1000
        &2.3156&2.6966&3.0315\\
LSUB2(2)&0.4389&0.7689&1.1980&1.3684&1.1115&--0.3116
        &--6.3126&     &\\
LSUB2(3)&0.4389&0.7703&1.2319&1.5567&1.8019&1.9821
        &2.1017&2.1670&2.0078\\
LSUB2(4)&0.4389&0.7702&1.2320&1.5615&1.8243&2.0409
        &2.2184&2.4663&2.5844\\
LSUB2(6)&0.4389&0.7702&1.2322&1.5637&1.8343&2.0692
        &2.2798&2.6501&2.9714\\
LSUB2(8)&0.4389&0.7702&1.2322&1.5637&1.8345&2.0698
        &2.2811&2.6540&2.9796\\
LSUB2(10)&0.4389&0.7702&1.2322&1.5638&1.8345&2.0700
        &2.2815&2.6557&2.9841\\
PT4(S)  &0.4389&0.7732&1.3708&2.6273&5.9333&13.236
        &27.038&86.933&\\
PT(W)   &0.5744&0.8624&1.2697&1.5822&1.8457&2.0778
        &2.2877&2.6603&2.9886\\
\noalign{\smallskip}\hline
\end{tabular*}
\end{table*}

\begin{table*}[t]
\setlength{\tabcolsep}{2.34mm}
% \newlength{\digitwidth} \settowidth{\digitwidth}{\rm 0}
\caption{Ground-state energy per plaquette at several values of
$\lambda$ for the 2D U(1) model. Shown are the results from the CCM
LSUB2($n$) scheme, and from the strong- and weak-coupling expansion
series, denoted as PT$m$(S) ($m$th order) and PT(W) respectively.
Also
shown are the results from other techniques as explained in the
text.}
\begin{tabular*}{\textwidth}{llllllllll}\hline\noalign{\smallskip}
            &\multicolumn{9}{c}{$\lambda$}\\
            \noalign{\smallskip}\cline{2-10}\noalign{\smallskip}
\multicolumn{1}{c}{Method}
            &\multicolumn{1}{c}{0.5}&\multicolumn{1}{c}{1}
            &\multicolumn{1}{c}{2}&\multicolumn{1}{c}{3}
            &\multicolumn{1}{c}{4}&\multicolumn{1}{c}{5}
            &\multicolumn{1}{c}{6}&\multicolumn{1}{c}{8}
            &\multicolumn{1}{c}{9}\\
            \noalign{\smallskip}\hline\noalign{\smallskip}
SUB1   &0.4391&0.7724&1.2430&1.5828&1.8597&2.1000&2.3156
              &2.6966&2.8686\\
LSUB2(2)&0.4386&0.7652&1.1468&1.1280&0.3019&--2.833&--15.11
              &    &\\
LSUB2(3)&0.4387&0.7681&1.2216&1.5371&1.7687&1.9265&2.0110
        &1.9584&1.8221\\
LSUB2(4)&0.4387&0.7681&1.2214&1.5428&1.7994&2.0123&2.1901
        &2.4585&2.5568\\
LSUB2(5)&0.4387&0.7681&1.2216&1.5442&1.8043&2.0237&2.2105
        &2.4977&2.6001\\
LSUB2(6)&0.4387&0.7681&1.2217&1.5453&1.8100&2.0407&2.2488
        &2.6207&2.7915\\
LSUB2(8)&0.4387&0.7681&1.2217&1.5454&1.8100&2.0404&2.2477
        &2.6142&2.7797\\
LSUB2(10)&0.4387&0.7681&1.2217&1.5454&1.8100&2.0405&2.2480
        &2.6155&2.7816\\
CBF     &0.4387&0.7677&1.2167&1.5335&1.7929&2.0201&2.2255&    &\\
HOZ     &      &      &1.215 &      &1.785 &      &2.2   &    &\\
PT4(S)  &0.4387&0.7690&1.3042&2.2898&4.8667&10.632&21.638&    &\\
PT8(S)  &0.4387&0.7673&1.1358&--0.738&--20.87&    &      &    &\\
Morningstar&   &0.7675&      &      &1.796 &      &      &&2.763\\
PT(W)   &0.5627&0.8434&1.2402&1.5447&1.8015&2.0276&2.2321
        &2.5917&2.7596\\
\noalign{\smallskip}\hline
\end{tabular*}
\end{table*}

It is worth mentioning that in quantum chemistry and other many-body
systems, the more relevant physical quantity is the so called
correlation energy which is defined as the difference between the
mean-field one-body (SUB1 in the present case) and the exact
ground-state energies. This correlation energy can be measured
experimentally in the cases of atoms and molecules. From Table~1 and 2
and Fig. 1, one can see that the correlation energy within the LSUB2
approximation (i.e., the LSUB2 results minus the SUB1 results) in the
$U(1)$ model is quite small, and much smaller than the total energy.
We suspect that this is true for lattice gauge field theories in
general. This is quite similar to the case in quantum chemistry where
the correlation energy is typically only a few percent at most of the
total energy. It is clear that a powerful many-body technique is
required in order to obtain a sensible numerical value for this
correlation energy.

The perturbation series in the weak-coupling limit ($\lambda
\rightarrow \infty$) is given by
% \buildrel{\lambda\rightarrow\infty}\under
\begin{equation}
 {E_g\over N_p}\rightarrow
      C_0\sqrt{\lambda}-{1\over8}C^2_0+{\cal O}(\lambda^{-1/2}),
\end{equation}
where $C_0 =1,\ 0.9833,\ 0.9581$ in 0D (i.e., one-plaquette or Mathieu
problem), 1D and 2D respectively. We also show the results from this
weak-coupling series in Table 1 and 2, denoted as PT(W). Although our
LSUB2($n$) schemes do not produce exactly these numbers, they do give
good results even for very large values of $\lambda$, as can be seen
from Table 1 and 2. From those results at large $\lambda$, we obtain,
by least squares fit, $C_0 \approx 1.0004, 0.9840, 0.9677$ in 0D,
1D, and 2D respectively.

\section{THE $U(1)$ MODEL IN 3D}

As discussed in Sec. 2, due to the geometrical constraints of the
Bianchi identity of Eq. (9), we have to employ the link variables
$\{A_l\}$ instead of the plaquette variables $\{B_p\}$ when taking
inner products for the 3D model. Since we are working in the gauge
invariant sector, the exact ground state $\vert\Psi_g\rangle$ should
be expressible by the plaquette variables $\{B_p\}$ alone.  Therefore,
we still write the 3D correlation operator $S$ and the ground-state
wavefunction $\vert\Psi_g\rangle$ in the same form as Eqs.  (16)-(18)
of the 1D and 2D cases. However, the inner products of Eqs.  (19)-(21)
should now represent integrals over all link variables $\{A_l\}$, as
defined by Eq. (5), namely
\begin{equation}
  \langle\Phi\vert\cos nB_p{\rm e}^{-S}H{\rm e}^S
   \vert\Phi\rangle_A =  E_g\delta_{n,0};\ \ \ \ n=0,1,2,...,
\end{equation}
for the energy-equation and one-plaquette equation, and
\begin{eqnarray}
  \langle\Phi\vert\cos n_1B_{p_1}\cos n_2B_{p_2}
   {\rm e}^{-S}H{\rm e}^S\vert\Phi\rangle_A &=& 0,\\
 \langle\Phi\vert\sin n_1B_{p_1}\sin n_2B_{p_2}
  {\rm e}^{-S}H{\rm e}^S\vert\Phi\rangle_A  &=& 0,
\end{eqnarray}
for the two-plaquette coefficients, where as before $n_1,n_2 =
1,2,...$ and $p_1 \not= p_2$. In the above equations, the plaquette
variables $\{B_p\}$ should be substituted by Eq. (2) before
integration, and the notation $\langle\cdots\rangle_A$ is defined by
the Eq. (5).

We again consider the LSUB2($n$) scheme which has been employed in the
1D and 2D models above. For the 3D model, it is clear that we need to
include the two perpendicular nearest-neighbour plaquette
configurations, as well as the two parallel in-plane ones which are
the only nearest-neighbour configurations in 1D and 2D. When
evaluating the integrals of Eqs. (26)-(28) over the link variables
$\{A_l\}$ after substitution of Eq.  (2) for all $\{B_p\}$, it is very
convenient to use exponential representations of the trigonometric
functions, namely
\begin{equation}
  \cos x = {1\over2} \left({\rm e}^{ix} + {\rm e}^{-ix}\right),
  \ \ \ \
  \sin x = {1\over2i} \left({\rm e}^{ix} - {\rm e}^{-ix}\right).
\end{equation}
We note that for the 1D and 2D cases, the integrals implicit in Eqs.
(26)-(27) yield results identical to Eqs. (19)-(21). This is not
surprising because the Bianchi identity of Eq. (9) is automatically
satisfied in 1D and 2D. (However, in 3D, the two sets of integrals
yield results which differ in the two-plaquette equations.)

As to the 1D and 2D cases, we find that the LSUB2(2) scheme reproduces
the 3D strong-coupling perturbation expansion up to fourth order,
\begin{equation}
  {E_g\over N_p} \rightarrow \lambda -{1\over4}\lambda^2
  + {3\over1280}\lambda^4 + {\cal O}(\lambda^6), \ \ \ \
  \lambda\rightarrow 0.
\end{equation}

\begin{table*}[t]
\setlength{\tabcolsep}{2.6mm}
% \newlength{\digitwidth} \settowidth{\digitwidth}{\rm 0}
\caption{Ground-state energy per plaquette at several values of
$\lambda$ for the 3D U(1) model. Shown are the results from the CCM
LSUB2($n$) scheme, and from the strong- and weak expansion series,
denoted as PT$m$(S) ($m$th order) and PT(W) respectively. Also shown
are the results from the Monte Carlo calculations of Hamer and Aydin
(MC) in Ref. [9] and the loop calculus (LC) of Ref. [8].}
\begin{tabular*}{\textwidth}{llllllllll}\hline\noalign{\smallskip}
            &\multicolumn{9}{c}{$\lambda$}\\
            \noalign{\smallskip}\cline{2-10}\noalign{\smallskip}
\multicolumn{1}{c}{Method}
            &\multicolumn{1}{c}{0.2}&\multicolumn{1}{c}{0.4}
            &\multicolumn{1}{c}{0.6}&\multicolumn{1}{c}{0.8}
            &\multicolumn{1}{c}{1.0}&\multicolumn{1}{c}{1.5}
            &\multicolumn{1}{c}{2.0}&\multicolumn{1}{c}{3.0}
            &\multicolumn{1}{c}{4.0}\\
            \noalign{\smallskip}\hline\noalign{\smallskip}
SUB1   &0.1900&0.3607&0.5133&0.6498&0.7724&1.0316&1.2430
          &1.5828&1.8597\\
LSUB2(2)&0.1900&0.3600&0.5100&0.6386&0.7435&0.8232&  &  &\\
LSUB2(3)&0.1900&0.3601&0.5105&0.6422&0.7563&0.9776&1.1316
          &1.3193&1.4097\\
LSUB2(4)&0.1900&0.3601&0.5105&0.6421&0.7561&0.9756&1.1230
          &1.2761&1.3027\\
LSUB2(5)&0.1900&0.3601&0.5105&0.6421&0.7561&0.9756&1.1227
          &1.2690&1.2617\\
LSUB2(6)&0.1900&0.3601&0.5105&0.6421&0.7561&0.9756&1.1230
          &1.2737&1.2885\\
LSUB2(8)&0.1900&0.3601&0.5105&0.6421&0.7561&0.9756&1.1230
          &1.2735&1.2868\\
MC      &0.1900&0.3600&0.5115&0.6203&      &      &      & &\\
LC      &0.1900&0.360 &0.51  &0.62  &0.71  &      &      & &\\
PT4(S)  &0.1900&0.3601&0.5103&0.6410&0.7523&0.9494&1.0375
        &0.9398&0.6000\\
PT(W)   &0.2768&0.4242&0.5373&0.6327&0.7167&0.8956&1.0464
        &1.2994&1.5126\\
\noalign{\smallskip}\hline
\end{tabular*}
\end{table*}

In Table 3, we show our numerical results for the ground-state energy
as a function of $\lambda$ for the LSUB2($n$) scheme for several
values of $n$ up to $n=8$. For comparison, we have also included
values obtained from the fourth-order strong-coupling perturbation
expansion (30) and from the weak-coupling perturbation theory
expression of Eq.  (25) with $C_0 = 0.7959$ as obtained from the Monte
Carlo calculations of Chin, Negele, and Koonin \cite{MC}. We also
include values obtained by Hamer and Aydin using a Monte Carlo method
(MC) \cite{MC}, and by Aroca and Fort using a loop calculus (LC)
\cite{loop}. From Table 3, we see that the LSUB2($n$) results converge
very well at low values of $\lambda$ ($\lambda \le 2$). They agree
well with Monte Carlo and loop calculus results for $\lambda < 0.8$.
However, the scheme seems to break down badly in the weak-coupling
regime ($\lambda >3$).  This is quite different from the results of
the similar LSUB2($n$) scheme in the 1D and 2D cases, where the CCM
results are still very good well into the weak-coupling regime
($\lambda > 10$).  We suspect that this difference is most likely the
result of the deconfining phase transition (probably second order) in
the 3D $U(1)$ model, which is predicted to occur at $\lambda
\approx 0.65$. In order to investigate this possibility, we have
also calculated within the LSUB2($n$) scheme the ``specific heat"
which is defined as the second-order derivative of the ground-state
energy per plaquette with respect to the coupling parameter $\lambda$.
Unfortunately, the specific heat results do not show any indication of
a phase transition. It is clear that the physical properties near the
phase transition are beyond the present low level approximation
scheme.

\section{CONCLUSION}

In this article, we have reviewed our application of the systematic
CCM approach to $U(1)$ lattice gauge models in various dimensions. For
the 1D and 2D models, we employ the plaquette variables $\{B_p\}$
which are the natural choice for the independent variables in these
cases, but in 3D we employ the link variables $\{A_l\}$ due to the
geometrical constraints. Our results from a local approximation scheme
reproduce the strong-coupling expansion series up to the fourth order
for all models considered. Furthermore, the LSUB2 results for the 1D
and 2D models are also quite reliable for $\lambda$ well into the
weak-coupling regime. We therefore conclude that the CCM comprises, in
effect, a well-defined analytical continuation or resummation of the
strong-coupling perturbation series, within the context of a natural
and consistent hierarchy.

Preliminary work on the low-lying excitation gap (glueball mass) and
on the non-Abelian $SU(2)$ model has also been carried out within the
pure gauge sector \cite{CCM}. In the same context, we believe that our
above formalism also provides a systematic approach to other
interesting physical quantities such as the string tension. The
generalization of our formalism from the pure gauge sector to the
charged sector can also be done in principle by including in sums of
Eq. (16) for the correlation operator $S$ not only terms corresponding
to closed paths (Wilson loops) on the lattice, but also terms
representing open paths corresponding tubes of electric
flux between staggered fermions.

The quality of the LSUB2 results for the 3D model in the weak-coupling
region ($\lambda >3$), however, is quite poor. This may reflect the
fact that the 3D $U(1)$ lattice gauge system experiences a deconfining
phase transition at a critical coupling $\lambda_c$. A possible
solution may be to look at improving the reference model state used in
the CCM. The electric vacuum state (constant state) which we have used
is the simplest possible, and could certainly bear improvement. One
option is to consider the use of a mean-field type state which
includes only one-body correlations, but already produces much better
results in the weak-coupling regime than the electric vacuum state.
Furthermore, our past experience for quantum spin lattice systems
clearly reveals that one has to go to high-order calculations of the
CCM in order to see possible phase transitions in the quantum systems
\cite{CCMspin2}. We believe that similar high-order approximations
should be able to reveal the critical properties of the deconfining
phase transition in the 3D model.

\vspace{2 \baselineskip}
\leftline{\large\bf Acknowledgements}
\vspace{\baselineskip}
We are grateful to S.A. Chin for many useful discussions, particularly
with regards to the geometrical constraints of the lattice gauge
models in 3D. One of us (R.F.B.) acknowledges the support of a research
grant from the Science and Engineering Research Council (SERC) of
Great Britain.

\end{document}